\documentclass[aps,pra,showpacs,twocolumn]{revtex4}%,preprint
\usepackage{graphicx}
\usepackage{amsmath}

\def\<{\langle}
\def\>{\rangle}

\begin{document}

\title{Secure Reusable Base-String in Quantum Key Distribution}
\author{Kai Wen$^1$, Fu Guo Deng$^{1,2}$ and Gui Lu Long$^{1,3}$}
\affiliation{$^1$ Key Laboratory for Atomic and Molecular NanoSciences and
Department of Physics, Tsinghua University, Beijing 100084, China; \\
$^2$   Institute of Low Energy Nuclear Physics, Beijing Normal University,
Beijing 100875,  China\\
 $^3$ Tsinghua National Laboratory for Information Science and
Technology, Tsinghua University, Beijing 100084 }
\date{June 26, 2007}

\begin{abstract}
Protecting secure random key from eavesdropping in quantum key distribution protocols
has been well developed. In this letter, we further study how to detect and eliminate
eavesdropping on the random base string in such protocols. The correlation between the
base string and the key enables Alice and Bob to use specific privacy amplification to
distill and reuse the previously shared base string with unconditional security and
high efficiency. The analysis of the unconditional secure reusable base string brings
about new concept and protocol design technique.
\end{abstract}

\maketitle

Quantum cryptography has received wide attention from people who pursue perfectly
secure communication. Since the first quantum key distribution (QKD) protocol by
Bennett and Brassard \cite{bb84}, the BB84 QKD protocol, scientists have developed
various techniques to improve its security and efficiency. Recently, unconditional
security of BB84 QKD protocol have been achieved  \cite{lc,mayers,sp}.

Most QKD protocols focus on detecting and eliminating eavesdropping on the random key
by encoding it to the qubits with different bases. The choices of the bases, namely the
base string likes an encryption key. Especially in some recent works, the base string
are shared before the protocol \cite{pab,pabshor,pabwg,core,qss,cpl}. So it is
interesting to investigate how to protect such encryption key from eavesdropping.

In this Letter, we explicitly prove how much unconditional secure base string
can be reused for the first time. The result not only increases the efficiency
of quantum cryptography, but also contributes to the foundation of quantum
information science and cryptography designs.

Suppose Alice and Bob share a common and random base string. Then Alice encodes the
qubits using this shared base string and sends them to Bob. Bob receives and measures
the qubits in the bases determined by the shared base string. Under ideal situation,
Alice and Bob can reuse the base string if no error is found in the check step. This
kind of sharing is able to greatly increase the efficiency of quantum key distribution,
because no qubit is wasted due to wrong choice of the measuring basis.

However, practically the channel errors are unavoidable. Eve may steal some information
about the secret base string. However, we will prove that Alice and Bob can estimate
and eliminate Eve's information on the base string by privacy amplification.  The proof
of our QKD protocol with shared and reusable base string is also based on the lemma
\cite{lc} that high fidelity implies low entropy and the similar reduction technique of
Shor and Preskill  \cite{sp}. But we extend the security analysis from one entangled
pair to a block of two entangled pairs. This extension gives correlation between the
error rates of the two entangled pairs.

We start with the entanglement distillation QKD protocol. Firstly, we can also
suppose that the secret and random base string is also generated from another
high-fidelity EPR pairs, denoted by the base pairs, namely, Alice and Bob both
measure their qubits of the base pairs respectively in the $Z$-basis. Secondly,
we let Alice and Bob postpone the measurements on the base pairs. Then Alice's
random Hadamard transformation on the second qubit of each communicating pair
is replaced by the controlled-Hadamard operation with her own qubit of one base
pair as the source qubit and the second qubit of one communicating pair as the
target qubit. Particularly, the control-Hadamard gate using the first qubit to
control the second qubit is denoted as $CH_{12}$.

\begin{figure}
\includegraphics[width=8cm]{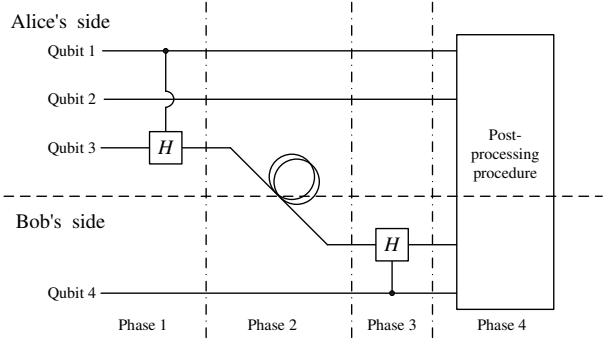}
\caption{Entanglement distillation QKD protocol with reusable base string in a block of
two EPR pairs}\label{edp_fig}
\end{figure}

Following the ideas above, we obtain the following protocol:
\newtheorem{protocol}{Protocol}
\begin{protocol}\label{edp}
Entanglement distillation QKD protocol with reusable shared base string
\end{protocol}
\begin{enumerate}
\item Alice and Bob share $2n$ base EPR pairs in the state $|\Phi^+\>^{\otimes 2n}$.
\item Alice prepares $2n$ communicating EPR pairs in the state $|\Phi^+\>^{\otimes
2n}$, and groups each communicating pair with one base pair to create $2n$ blocks. Fig.
\ref{edp_fig} shows the operations on one block in the protocol, in which the 1st and
the 4th qubits form a base pair and the 2nd and the 3rd qubits form a communicating
pair.

\item In each block, Alice applies $CH_{13}$, as shown in phase 1 of
Fig. \ref{edp_fig}.

\item Alice sends the 3rd qubit of each block to Bob, as shown in phase 2 of
Fig. \ref{edp_fig}.

\item Bob receives the qubits. In each block, he applies $CH_{43}$, as shown
in phase 3 of Fig. \ref{edp_fig}. Then he publicly announces the reception.

\item The following steps are post-processing procedure. Alice and Bob randomly
permute the blocks and agree on $n$ random blocks out of the $2n$ blocks as
check blocks.

\item In each check block, Alice and Bob both measure their own qubits of the
communicating pairs in $Z$-basis, and publicly compare their results to obtain
the channel bit error rate $e$. If there are too many errors, they abort the
protocol.

\item By estimating the bit error rate on the communicating pairs in the code
blocks from the checking process, Alice and Bob apply an entanglement
purification protocol(EPP) to distill $m$ communicating pairs with high
fidelity from the $n$ corrupted communicating pairs. Then they measure them
both in $Z$-basis to establish an $m$-bit secret key.

\item Alice and Bob can also estimate the phase error rate on the base pairs in
all blocks as not greater than $2e$. Then Alice and Bob apply another EPP to
distill $2m'$ base pairs with high fidelity from the $2n$ corrupted base pairs.
Then they measure them both in $Z$-basis to establish a $2m'$-bit secret base
string.
\end{enumerate}

The key point to prove the unconditional security of Protocol \ref{edp} is the
estimation of the error rates on the two kinds of the EPR pairs, shown in step 8 and 9.
When the channel is noisy, we suppose without loss of generality the errors of the
channel all come from Eve's manipulation of the quantum system. Assuming that Eve can
perform arbitrary coherent attack on all blocks. Here we apply quantum de Finetti
representation  \cite{deFinetti}, and we only consider the asymptotic situation of
large $n$. Thanks to the random permutation of the blocks in step 6 of Protocol
\ref{edp}, the $2n$ blocks are exchangeable and satisfy the condition of quantum de
Finetti representation. Therefore, the final state of the total $2n$ blocks is a
mixture of product state, namely, the density matrix of the final state is
\begin{eqnarray}
\rho'_{all} = \int p_{\rho'} \rho'^{\otimes 2n} d\rho',\label{deFi}
\end{eqnarray}
in which $\rho'$ is chosen from any possible corrupted density matrix of one
block and $p_{\rho'}$ is its weight. Due to the linear sum of different
$\rho'^{\otimes 2n}$ in the final density matrix, the results of any
measurement on $\rho'_{all}$ are also the linear weighted sum of measurement
results on different $\rho'^{\otimes 2n}$. So we can restrict our analysis
within one possible value of $\rho'$.

Considering the case with one possible value of $\rho'$, we find that the results of
measuring any operator on each block of the final state are effectively an independent
and identical distribution. The bit and phase error rates of the base pairs, denoted as
$E^{bit}_{base}$ and $E^{ph}_{base}$, and the bit and phase error rates of the
communicating pairs, denoted as $E^{bit}_{comm}$ and $E^{ph}_{comm}$,  are the rates of
obtaining $-1$ when measuring $Z_1 Z_4$, $X_1 X_4$, $Z_2 Z_3$ and $X_2 X_3$
respectively in the blocks. Because all these four measurement operators are commute to
each other, we are able to apply classical probability here. By the central limit
theorem, all these error rates are equal to the expected values of measurement results
of corresponding operators in one block, with very large probabilities. As a result, we
can only analyze the error rates of one block in one possible state of $\rho'$.

When we only study one block, the initial state is $\rho_0$ defined as
\begin{eqnarray}
\rho_0 = |\psi_0\>\<\psi_0|\label{initial_state},
\end{eqnarray}
where $|\psi_0\> = \frac{1}{2}(|00\>+|11\>)_{14}\otimes(|00\>+|11\>)_{23}$. Any
possible final state, $\rho'$, is obtained by arbitrary operation of Eve. A general
operation of Eve can be described by a superoperator on the 3rd qubit in each block
\cite{3state}, which is transmitted via the channel. We denote the superoperator as
$\S$. Then from the protocol, we obtain that
\begin{eqnarray}
\rho' = CH_{43} \S( CH_{13} \rho_0 CH_{13}^\dag ) CH_{43}^\dag.
\end{eqnarray}
A general superoperator of Eve's operation is described as
\begin{eqnarray}
\S(\rho_1) = \sum_{\mu} M_\mu \rho_1 M_\mu^\dag,\label{sp}
\end{eqnarray}
in which $\rho_1 = CH_{13} \rho_1 CH_{13}^\dag$ and $M_\mu$ is an arbitrary
matrix acting on the 3rd qubit, namely,
\begin{eqnarray}
M_\mu = \left(\begin{array}{cc} a^\mu_{11} & a^\mu_{12} \\ a^\mu_{21} &
a^\mu_{22} \end{array}\right)_3.
\end{eqnarray}

Due to the linearity of the superoperator, we first calculate the error rates on each
$M_\mu$. The results are
\begin{eqnarray}
E^{bit,\mu}_{comm} &=& \frac{1}{8} ( |a^\mu_{11}-a^\mu_{22}|^2+
|a^\mu_{12}-a^\mu_{21}|^2 +\nonumber \\
&& 2 |a^\mu_{12}|^2+ 2|a^\mu_{21}|^2 ), \\
E^{ph,\mu}_{comm} &=& E^{bit,\mu}_{comm},\\
E^{bit,\mu}_{base} &=& 0,\\
E^{ph,\mu}_{base} &=& E^{bit,\mu}_{comm}+\nonumber \\
&& \frac{1}{8}\{[-a^\mu_{11}+a^\mu_{22}] [(a^\mu_{12})^*+(a^\mu_{21})^*] + \nonumber\\
&&[-(a^\mu_{11})^*+(a^\mu_{22})^*] [a^\mu_{12}+a^\mu_{21}] \}\\
&\leq&  2 E^{bit,\mu}_{comm}..
\end{eqnarray}

Therefore, by using the linearity of Eq. (\ref{deFi}) and (\ref{sp}), we obtain the
relationship between the four kinds of error rates, namely,
\begin{eqnarray}
E^{ph}_{comm} &=& E^{bit}_{comm}, \\
E^{bit}_{base} &=& 0, \\
E^{ph}_{base} &\leq& 2 E^{bit}_{comm}.
\end{eqnarray}
To interpret the above results, we note that the controlled-Hadamard operations
serves as the random Hadamard operations, which make the bit and phase error
rates of the communicating pairs symmetric. As no qubit of the base pairs are
transmitted through the channel, their bit error rate is 0, while their phase
error rate is the result of the propagation of errors on the communicating
pairs.

Moreover, because Alice and Bob only need to know $E^{bit}_{comm}$, $E^{bit}_{comm}$
can be best estimated from the comparison of the $Z$-basis measurement results of $n$
check communicating pairs, in step 7 of Protocol \ref{edp}, as well as the channel bit
error rate $e$. Thus knowing the bounds of the error rates, Alice and Bob can employ
two EPPs on the communicating and base pairs respectively, shown in step 8 and 9. If
both EPP are successfully, they will distill both high-fidelity base and communicating
EPR pairs which implies low entropy of Eve's information  \cite{lc}. Therefore, not
only secret shared key is established but also secret base string can be reused in the
future, after $Z$-basis measurements on these pairs. Note that no base pairs are
sacrificed in the checking process so total $2n$ base pairs can be used in EPP.

So far, we have shown the unconditional security of Protocol \ref{edp}. Our final goal
is to derive a prepare-and-measure protocol with reusable shared base string from
Protocol \ref{edp}. The reduction lies on the fact that some of the final $Z$-basis
measurements in step 8 and 9 commute to other operations and measurements in the
protocol, and that can be brought forward to the beginning of the protocol without
affecting the security  \cite{sp}.

Firstly, both Alice and Bob's final measurements on both pairs can be brought before
the error correcting procedure in step 8 and 9. If we use EPP with one-way classical
communications, the result effectively changes the EPP to error correction with
Calderbank-Shor-Steane(CSS) codes  \cite{sp} on single qubits. A CSS Code $Q(C_1, C_2)$
employs two classical linear code $C_1$ and $C_2$, in which $C_1$ and $C_2$ are used
for correcting bit and phase errors respectively and $C_2 \subset C_1$  \cite{css}.
Moreover, because the bit error rate on the base pairs is always $0$, we only need
$C_2$ to correct the base pairs.

Secondly, Alice's final measurements commute with the controlled-Hadamard gates
applied in step 3. It can be also verified that Bob's final measurements on the
base pairs commute to the gates in step 5.

Thirdly, no measurement of the phase error rates are required, because we have
proved that the upper bounds of the two kinds of phase error rates can be
estimated using the channel bit error rate $e$. Thus Alice and Bob's final
measurements commute to the measurements in step 7.

Finally, we have successfully brought Alice's final measurements to the
beginning of Protocol \ref{edp}. We are also able to bring Bob's final
measurements on the base pairs to the beginning of the protocol, and to bring
those on the communicating pairs to Bob's reception of the qubits. We summarize
the result of the equivalent transformations as the following protocol.

\begin{protocol}\label{bb84}
BB84 QKD protocol with reusable shared base string
\end{protocol}
\begin{enumerate}
\item Alice and Bob share $2n$-bit binary base string $b$.

\item Alice prepares $2n$ qubits. If the corresponding bit value of $b$ is $0$,
she randomly prepares the qubit in $|0\>$ or $|1\>$; otherwise she randomly
prepares the qubit in $|+\>$ or $|-\>$.

\item Alice sends the qubits to Bob.

\item Bob receives the qubits and immediately measure them in certain basis according
his $b$. Then he publicly announces the reception.

\item Alice and Bob agree on $n$ qubits as check qubits. The rest $n$ qubits are code qubits.

\item Alice and Bob publicly compare the bit values on the check qubits and obtain
the channel bit error rate $e$. If there are too many errors, they abort the
protocol.

\item Alice and Bob select a CSS code $Q(C_1, C_2)$ that are capable of correcting
both bit and phase errors on the code qubits, which are both $e$. They employ
$C_1$ to correct the bit errors in the measurement results of the code qubits.
Then they use the cosset of the corrected results to $C_2$ as the final key.

\item Alice and Bob select another linear code $C'_2$ that are capable of
correcting the phase errors on the hypothetic base EPR pairs represented by
$b$, whose rate is at most $2e$. They use the cosset of $b$ to $C'_2$ as final
reusable secret base string.
\end{enumerate}

\begin{figure}
\includegraphics[width=6cm]{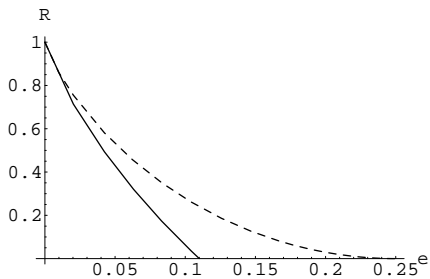}
\caption{The generation rates of final secure key(solid line) and base
string(dashed line)}\label{rate}
\end{figure}

Now we analyze the rates of generating random key and reusable base string. The maximal
achievable generation rate of the final secure key  \cite{sp,gllp} is
\begin{eqnarray}
R_k(e) = 1 - H(E^{bit}_{comm}) - H(E^{ph}_{comm}) = 1 - 2H(e),
\end{eqnarray}
in which $H(x)= -x \log_2 x - (1-x) \log_2 (1-x)$. The length of the final key is $n
R_k$. Similarly, if $2e < 0.5$, the maximal achievable generation rate of the final
reusable base string  \cite{sp} is
\begin{eqnarray}
R_b(e) = 1 - H(E^{bit}_{base}) - H(E^{ph}_{base}) \leq 1 - H(2e).
\end{eqnarray}
The length of the final reusable base string is $2n R_b$. We plot these two
rates on Fig. \ref{rate}. We find that the maximal error rate that gives
non-zero generation rate of the base string is $25\%$, much larger than that of
the key of about $11\%$.

\begin{figure}
\includegraphics[width=6cm]{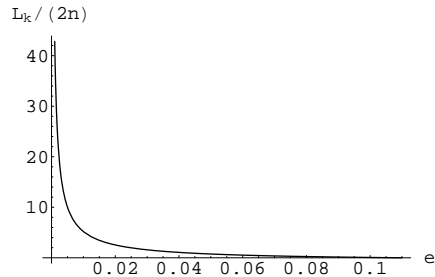}
\caption{$L_k / (2n)$ verse the channel bit error rate $e$}\label{totalrate}
\end{figure}

Suppose Alice and Bob initially share a $2n$-bit base string, they use it to
encode $2n$ qubits in Protocol \ref{bb84}. After the error correction and
privacy amplification, they get $n R_k(e_1)$-bit key and the $2n R_b(e_1)$-bit
base string remains. Then they use the base string again to encode $2n
R_b(e_1)$ qubits in the next protocol and obtain another $n R_b(e_1)
R_k(e_2)$-bit key; the remaining base string becomes $2n R_b(e_1) R_b(e_2)$. In
this way, they repeat Protocol \ref{bb84} again and again. If the channel bit
error rate does not change, the total length of key generated from the initial
$2n$-bit base string is
\begin{eqnarray}
L_k &=& n R_k(e)(1 + R_b(e) + R_b(e)^2 + \cdots) = \frac{n R_k(e)}{1 -
R_b(e)}.\nonumber\\
\end{eqnarray}
We also plot $L_k/(2n)$ on Fig. \ref{totalrate} and find that if the channel bit error
rate is low enough, a small length of initial base string can generate much longer
random key.

In conclusion, we have shown that if Alice and Bob first share a secret base string and
use them in BB84 QKD protocol to encode the qubits, they are able to reuse this shared
information in the future by the distillation of the base string using certain privacy
amplification methods. In particular, in the part of the qubits successfully received
by Bob, the generation rate of the base distillation is also related to the channel bit
error rate, and higher than that of the key distillation. Furthermore, as the bit error
rate of the base string is zero, we need only use a classical linear code instead of
CSS codes to distill the base string, so the distillation is much simpler and more
efficient. Secure reusable base string is contrasting different, and it may lead new
strategy in protocol designs in quantum cryptography, hence contributes to the
foundation of quantum information science.

The authors thank Kiyoshi Tamaki for helpful discussions. This work is
supported by the National Fundamental Research Program Grant No. 2006CB921106,
China National Natural Science Foundation Grant Nos. 10325521, 60433050, the
SRFDP program of Education Ministry of China, No. 20060003048.

\end{document}